\documentstyle[aps,prl,epsfig,floats]{revtex}
\begin{document}

\twocolumn[\hsize\textwidth\columnwidth\hsize\csname %
@twocolumnfalse\endcsname

\draft

\title{Impurity Scattering from $\delta$-layers in Giant
Magnetoresistance Systems}

\author{C.H. Marrows* and B.J. Hickey}
\address{Department of Physics and Astronomy, E.C. Stoner Laboratory,
University of Leeds, Leeds. LS2 9JT United Kingdom}
\date{\today}
\maketitle
\begin{abstract}
The properties of the archetypal Co/Cu giant magnetoresistance
(GMR) spin-valve structure have been modified by the insertion of
very thin (sub-monolayer) $\delta$-layers of various elements at
different points within the Co layers, and at the Co/Cu interface.
Different effects are observed depending on the nature of the
impurity, its position within the periodic table, and its location
within the spin-valve. The GMR can be strongly enhanced or
suppressed for various specific combinations of these parameters,
giving insight into the microscopic mechanisms giving rise to the
GMR.
\end{abstract}


]

Ever since the development of the first transistor, solid-state
science and technology has sought a proper description of the
details of electronic transport in heterostructures. The past few
years has seen a remarkably high level of activity in the area of
magnetic heterostructures on the nanometer scale, not least in the
area of giant magnetoresistance (GMR) \cite{GMRrev}, observed in
ultrathin layered structures featuring transition metal
ferromagnets that can have the relative orientation of their layer
moments altered by a magnetic field. The broad physical picture
describing GMR is that it arises from spin-dependent scattering,
so that parallel or antiparallel magnetic layer moments correspond
to aligned or anti-aligned filters for spin-polarized current.
Approaches to the theory based on the Boltzmann formalism
\cite{c&b,dieny} can give a good phenomenological description of
the basic effects. Early quantum pictures used a
free-electron-like band, evaluating the Kubo formula for the case
of spin-dependent scattering potentials \cite{levy,vedyayev}. More
recent theoretical treatments have emphasized the importance of
the electronic structure to the GMR
\cite{bauer,butler1,butler2,tsymbal,sanvito}, which yield a better
quantitative agreement with experiment.

Nevertheless these theories only consider pairs of materials ({\it
e.g.} Fe/Cr, Co/Cu), limiting the understanding of more complex
experimental structures. One area of contention is the microscopic
location of the spin-dependent scattering -- in the bulk or at the
interface of the ferromagnetic layers. It has been attempted to
get directly at the microscopic origin of the GMR by deliberately
doping with impurities. This was reported for Fe/Cr multilayers
using a few different dopants placed at the
interface\cite{baumgart,gurney}. The different impurities have
different values for the scattering spin-asymmetry $\alpha$,
defined as the ratio of spin $\uparrow$ to spin $\downarrow$
scattering from the impurity $\rho_\downarrow / \rho_\uparrow$
\cite{vandenberg,c&f,alpha}, an essentially phenomenological
parameter -- only in the last few years have attempts been made to
determine $\alpha$ from from electronic band structure
calculations\cite{irkhin}. Similar interfacial doping experiments
were reported by Shinjo \cite{shinjo}. Nevertheless these
experiments were carried out using AF coupled superlattices,
complicating the interpretation, as the AF state is ill-defined,
and is easily degraded by the insertion of the dopants, leading to
a loss of GMR merely due to loss of AF alignment \cite{perez}.
Meanwhile Vouille {\it et al.} have studied the effects of doping
various elements into the magnetic layers as alloys\cite{vouille}
-- although this varied the $\alpha$ of the dopants, determining
the relative bulk and interface contributions of these scatterers
is model dependent.

A noteworthy theoretical treatment of the both the position and
spin asymmetry ($\alpha$) properties of impurities in Co/Cu
multilayers has been given by Zahn {\it et al.} \cite{zahn}. Using
the tight-binding Korringa-Kohn-Rostoker technique they were able
to calculate the local density of states \cite{binder} and hence
the effect of the impurity scatterers on the GMR. In this way
direct conclusions can be drawn about the relative importance of
bulk and interface scattering -- however these ideas have not been
tested at all stringently by any of the experiments cited above.

In this Letter we wish to address these issues, reporting on
experiments in which we have systematically doped archetypal Co/Cu
spin-valves by the insertion of $\delta$-layers of various
elements to localize scattering with a certain value of $\alpha$.
The use of spin-valves removes the difficulties in ensuring a
proper AF alignment, as we always have a clear distinction between
parallel ($\uparrow \uparrow$) and antiparallel ($\uparrow
\downarrow$) moment alignments -- so we can be certain to have
measured the full GMR amplitude, defined as $(\rho_{\uparrow
\downarrow} - \rho_{\uparrow \uparrow}) / \rho_{\uparrow
\uparrow}$. The previous experiments used only a few impurities,
we have prepared a much larger set of samples to systematically
study the dependence of the GMR on the changes in electronic
structure caused by the introduction of a wide variety of
different dopants. In addition our $\delta$-doping technique
yields important information on the position dependence of the
impurities that cannot be obtained by forming alloys or
interfacial layers alone. We have observed long ranged
interactions between several different impurities and the
interfacial spin-dependent scattering, over distances up to an
order of magnitude greater than those previously reported
\cite{parkin} or predicted \cite{gijs&bauer}.

\begin{figure*}
\centerline{\epsfig{figure=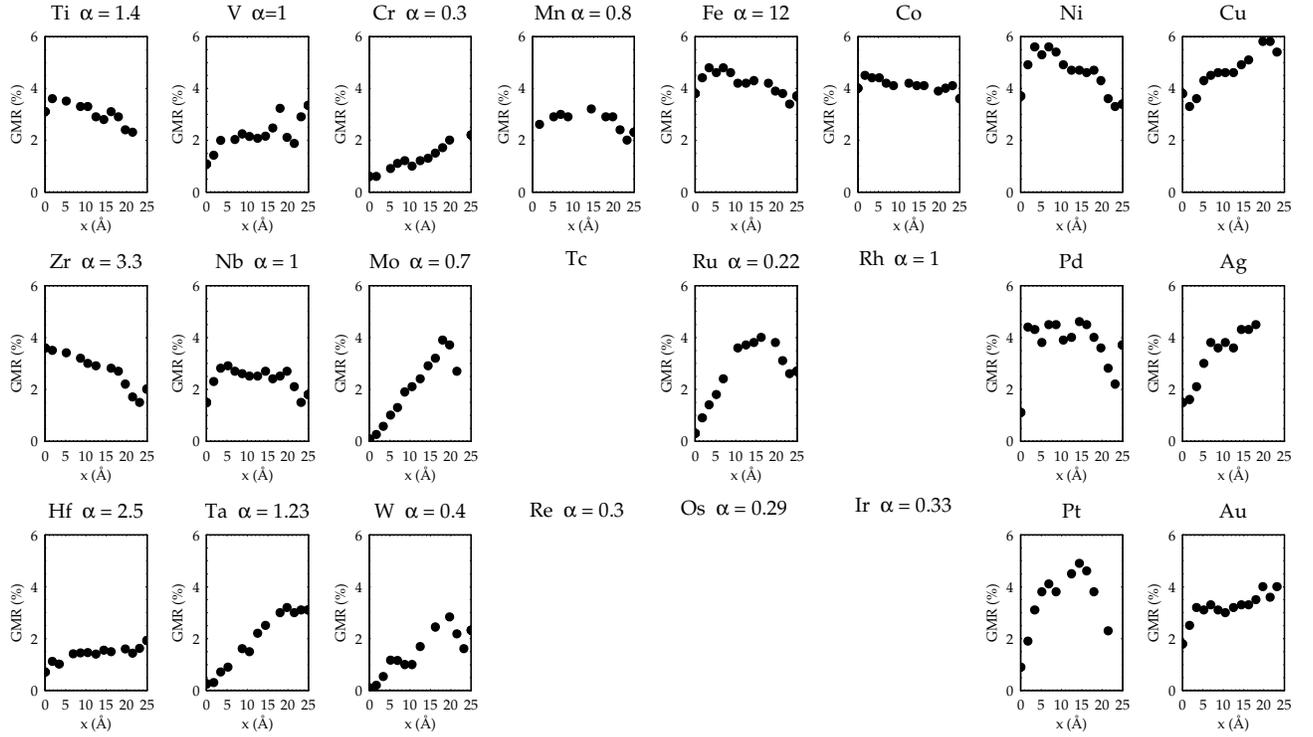,width=17.5cm}}\medskip
\caption{Position dependence of the giant magnetoresistance for
various transition metal impurities in the Co layer of the
spin-valves. Elements for which $\alpha$ values are available from
reference \protect \cite{vandenberg} have these values noted at
the top of each panel. The graph width represents the Co layer
thickness. As $x$ increases in each graph we move from the Co/Cu
interface to outermost surface of the Co layers.} \label{periodic}
\end{figure*}

%\begin{figure*}
%\centerline{\epsfig{figure=re.eps,width=12.5cm}}\medskip
%\caption{Position dependence the giant magnetoresistance for Al
%and selected 4$f$ impurities in the Co layer of the spin-valves.
%The graph width represents the Co layer thickness. As $x$
%increases in each graph we move from the Co/Cu interface to
%outermost surface of the Co layers.} \label{re}
%\end{figure*}

The samples were deposited by dc magnetron sputtering in a
computer controlled custom vacuum system with a base pressure of
$2\times 10 ^{-8}$ Torr. The substrates were pieces cut from (001)
Si wafer, the working gas was 3.0 mTorr of Ar, and typical
deposition rates were $\sim$3 \AA s$^{-1}$. The substrates are
heat-sunk during deposition so that the temperature does not rise
by more than a few $^\circ$C above ambient. Magnetoresistance was
measured by a standard dc 4-probe method, at room temperature. The
sample structure comprises those elements found in a typical
spin-valve - two Co layers separated by Cu, with an FeMn pinning
layer. The impurity $\delta$-layer is inserted into the Co at
different points so that the overall structure is as follows: Si
substrate / Ta(50) / Co(25-$x$) / X / Co($x$) / Cu(30) / Co($x$) /
X / Co(25-$x$) / FeMn(80) / Ta(25); all thicknesses are given in
\AA. Since both the Ta and FeMn have resistivities of much greater
than 100 $\mu\Omega$cm, we should expect most of the in-plane
conduction to take place in the GMR active Co/Cu/Co sandwich. In
all cases the amount of impurity corresponds to a few tenths of a
monolayer - we used standard conditions of 0.5 s deposition using
a power density of 1 W/cm$^2$. In some cases the introduction of
the $\delta$-layer close to the Co/FeMn interface reduced the
exchange bias to the point where the $\uparrow \downarrow$ state
cannot be accessed, such data points have been omitted from all
the Figures that we present.

Structural changes have been noted in similar experiments: the use
of sub-monolayer amounts of impurities as surfactants
\cite{miranda,egelhoff} can change the resistivity as they alter
growth modes whilst floating out of the film on the growth front.
We have tested for such effects and not found them: there is
little change in the observed GMR if we restrict ourselves to a
$\delta$-layer in only one or other Co layer. Since the $\delta$
layer is being moved, in sequential samples, in opposite
directions through the stack in these two cases the effects cannot
be due to changes due to its floating out, as surfactant effects
can only occur in layers deposited after the $\delta$-layer.

In Fig. \ref{periodic} the GMR is plotted against the position of
the dopant $\delta$-layer for a variety of elements from the
central part of the transition metal block. Firstly the reader
should note that the graph for Co is quite flat at $\sim$4.5\%,
and this can be regarded as the control experiment - a
$\delta$-layer of Co was inserted into both Co layers. This modest
value is due to the thinness of the Co layers compared to those
used in device applications \cite{devgmr}. It is immediately
evident that it is not possible to increase the GMR of a Co/Cu
structure by putting {\em any} other impurity at the interface,
previously thought to be that part of the structure most
susceptible to changes in chemical species \cite{parkin}.

On the other hand, certain impurities will increase the GMR when
placed {\em within} the Co, contrary to commonly held views about
the pre-eminence of the interfaces for GMR. The neighboring plots
for Ni and Fe show a similar behavior - both curves show a
pronounced rise in GMR when the $\delta$-layer is placed just
behind, but not at, the interface. The effect is larger for Ni --
almost 50\% higher. One possible interpretation is that the
$\delta$-layer forms a second highly effective spin-filter just
behind the first filter of the Co/Cu interface. Although no value
is reported in Ref. \onlinecite{vandenberg} for Ni, the value of
$\alpha_{\rm Fe}$=12 given for Fe in Co leads one to suppose that
the value of $\alpha_{\rm Ni}$ must also be $\gg 1$, and likely to
be even higher still than the value for Fe.

The effects of Cu impurities are also of particular interest. When
close to the interface there is little effect, or a small
suppression, due presumably to the artificial creation of a more
interdiffused, alloyed interfacial layer. However, once the Cu is
deep inside the layer we see an enhanced GMR, somewhat unexpected
in the light of the fact that these are non-magnetic impurities.
An obvious comparison here is with the large bulk spin-anisotropy
in the resistivity of Ni layers doped with Cu observed by Vouille
{\it et al.} \cite{vouille}.

Within the group of noble metals, the GMR is lower for Ag
impurities than for Cu, and lower still for Au. The behavior is
consistent with greater spin-orbit scattering -- the heavier
elements flip spins more readily, mixing the spin current
channels. A comparison with, for example, Pd and Pt is consistent;
the GMR recovers more rapidly as Pd is moved away from the Co/Cu
interface. Both these elements, with strong Stoner susceptibility
enhancements in the bulk, are readily polarized by the Co matrix,
leading to little loss in GMR.

On the other hand the graphs for Cr, Mo, Ru, Ta and W all show
that the insertion of the $\delta$-layer at the interface almost
totally suppresses the GMR. As the impurity is moved back into the
Co the GMR rises in a roughly linear fashion. For Ru and Ta the
GMR appears to plateau when the dopant is $\sim$10 and 20 \AA\
from the interface respectively. This is exactly the behavior
expected given the importance attached to interfacial scattering,
but the length scale is greater than that of only $\sim$2.5\AA\
previously reported when Co $\delta$-layers were inserted into
NiFe\cite{parkin}, suggesting that the lengthscales involved in
discussions of interfacial or bulk scattering must be highly
material system dependent. For all of these materials but Ta, the
reported $\alpha$ value is $<1$. The value of $\alpha_{\rm Ta} =
1.23$ appears to be an overestimate.

The data for Mn, V and Nb also look similar. These elements have
$\alpha$ values reported $\simeq 1$, and we see that the
dependence on the position of the dopant layer is quite weak. The
GMR is suppressed wherever the $\delta$-layer is placed.
%Although there are no values for $\alpha$ reported in the
%literature for impurities outside the 3$d$ block in a Co host, we
%find in the review of Campbell and Fert \cite{c&f} that
%$\alpha_{\rm Al}$= 1.7 in Ni, and 8.6 in Fe, so we might
%reasonably expect $\alpha_{\rm Al}>$1 in Co. Indeed, as can be
%seen in Fig. \ref{re}, the curve has the same characteristic shape
%as can be seen for Ti and Zr in Fig. \ref{periodic} ($\alpha_{\rm
%Ti}$=1.4, $\alpha_{\rm Zr}$=3.3).
There is little or no suppression of the GMR when the elements Ti
or Zr, both with $\alpha > 1$, are introduced into the interfacial
region of the Co layer. The effects of Hf are anomalous in this
regard, possibly either $\alpha_{\rm Hf}$=2.5 is an over-estimate,
or the high nuclear charge of Hf leads to a large spin-orbit
scattering term. This is to be compared with the results found for
Ta.

%On the other hand, 4$f$ impurities suppress the GMR at any point
%within the Co layer, as can be seen in Fig.\ref{re} - indicating
%that there is little spin-dependence in the scattering from these
%defect sites. We might also expect strong spin-flip scattering due
%to spin-orbit interactions with the partly filled 4$f$ shell, this
%mixing of the spin current channels will also reduce the GMR.
%These data bear comparison with those for Mn, V and Nb shown in
%Fig. \ref{periodic}.

\begin{figure}
\centerline{\epsfig{figure=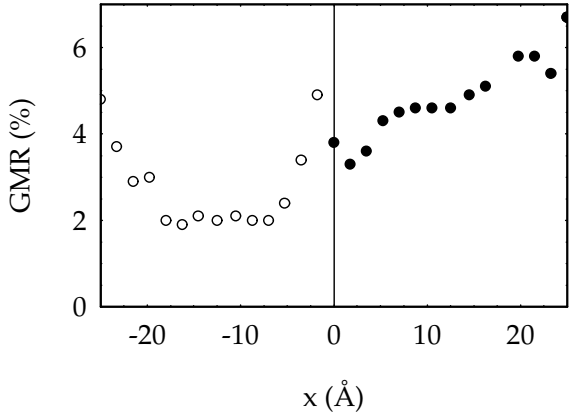,width=8cm}}\medskip
\caption{Dependence of the GMR on the position of Co impurities in
Cu (open symbols) or Cu impurities in Co (solid symbols). $x=0$
corresponds to the position of the Cu/Co interface.} \label{cocu}
\end{figure}

It is also of interest to pose the question regarding the effects
of impurities in the Cu spacer layer. The reader's attention is
drawn to Fig. \ref{cocu}, where the GMR of spin-valves with Co(Cu)
impurities in the Cu(Co) layer(s) is presented. The data for the
Cu impurities (solid symbols) is taken from Fig. \ref{periodic}.
As we have seen, the GMR rises as the Cu moves back into the Co
after a small suppression close to the interface. On the other
hand Co impurities in the Cu spacer strongly reduce the GMR with
only a weak position dependence unless they are close to the
interface. We should expect that Co atoms or clusters isolated in
the Cu should behave (super)paramagnetically, leading to
spin-independent scattering when averaging over time or position
in the film, as in practical measurements. The decay length of
$\sim$10 \AA\ is therefore a direct measure of the range of
significant exchange interactions for the Co impurities in Cu.
Further experiments with other impurities in the spacer layer are
all consistent with the same general picture - a position
independent suppression of the GMR due to a shortening of the mean
free path in the crucial spacer layer, unless the impurity is
within two or three atomic sites of the interfacial region, where
the impurity can begin to affect nature of the interfacial
scattering.

We find that the experimental results are at odds with the
published theoretical predictions of Zahn {\it et al.} \cite{zahn}
in the following important ways: impurities with $\alpha < 1$
suppress the GMR, usually to a great extent when at the interface,
and still have a considerable effect when several lattice
constants away from the interface; impurities with $\alpha > 1$
sometimes do provide an enhancement of the GMR, but it is only to
be found when they are a few \AA\ behind the Co/Cu interface; and
impurities in the spacer layer have a dramatic effect by lowering
the GMR. There are two omissions in the theory of Zahn {\it et
al.} which may lead to inaccurate predictions: a lack of interband
transitions, found to have an important effect on conductivity
calculations when realistic levels of disorder are included
\cite{tsymbal}; and vertex corrections are required for an
accurate description of impurity scattering \cite{swihart}.

The results of more sophisticated calculations by Binder {\it et
al.} \cite{binder2}, are qualitatively much more in accord with
the observations that we report here. Self-consistently calculated
impurity potentials were used, as well as a more correct
description of the microscopic transport processes including
state-dependent relaxation times and proper account taken of the
scattering-in term. In particular the predictions of the change in
GMR when moving the $\delta$-layers of specific materials from the
interface in to the bulk of the Co show remarkable similarities
with the observations and the sign of this change exhibits strong
correlations with the sign of the exchange interaction calculated
between the local moment of the impurity ion and the Co matrix.

The comprehensive nature of the data set allows some general
conclusions to be drawn -- there are consistent trends in the data
for impurity $\delta$-dopants with $\alpha <1$, $\simeq 1$, and $>
1$. The position dependence of the scattering that leads to the
GMR has been shown to be remarkably rich, and has important
implications for what is meant when bulk or interface scattering
is discussed. There are three rather striking results, deserving
of theoretical explanation: the opposite slopes of the data for
$\alpha>1$ or $<1$, as exemplified by Zr and Mo; the significant
increase in GMR caused by the insertion of 3$d$ ferromagnet
dopants {\em behind} the Co/Cu interface; and the marked increase
in GMR when a nonmagnetic impurity, Cu, is embedded deep in the
bulk of the Co. As well as suggesting possible routes to
optimizing GMR materials for devices, any theory found to be
capable of reproducing all these effects must contain the correct
physics of GMR at a deep level.

%\acknowledgements

We should like to express our gratitude to J. Binder, P. Zahn and
I. Mertig for thought-provoking discussions and access to
unpublished theoretical results. We should also like to thank
E.Yu. Tsymbal for helpful comments and suggestions. C.H. Marrows
would like to thank the Royal Commission for the Exhibition of
1851 for financial support.

% \begin{figure}
% \caption{}
% \label{}
% \end{figure}

% \begin{table}
% \caption{}
% \label{}
% \end{tabular}
% \end{table}


\begin{references}
\bibitem{alpha}We would wish to note that the values for the
spin-asymmetry parameter given in the previous two references show
considerable experimental scatter and moreover are for the
specific case of dilute impurities in a bulk host. However they
can at least be taken as a general guide.
\end{references}
\end{document}